\documentclass[preprint,12pt]{elsarticle}

\usepackage{amssymb}

\journal{Solid State Communications}

\begin{document}

\begin{frontmatter}



\title{Towards coherent optical control of a single hole spin: Rabi rotation of a trion conditional on the spin state of the hole}

\author[label1]{A.~J.~Ramsay\corref{cor6}}
\ead{a.j.ramsay@shef.ac.uk}
\author[label1]{S.~J.~Boyle}
\author[label1]{T.~M.~Godden}
\author[label1]{R.~S.~Kolodka}
\author[label1]{A.~F.~A.~Khatab}
\author[label2]{J.~B.~B.~Oliveira}
\author[label1]{J.~Skiba-Szymanska}
\author[label4,label5]{H.-Y.~Liu}
\author[label4]{M.~Hopkinson}
\author[label1]{A.~M.~Fox}
\author[label1]{M.~S.~Skolnick}

\address[label1]{Department of Physics and Astronomy, University of Sheffield, Sheffield S3 7RH, U.K.}
\address[label2]{Departamento de F$\mathrm{\acute{i}}$sica, Universidade Estadual Paulista - UNESP, Bauru-SP, 17.033-360, Brazil}
\address[label4]{Department of Electronic and Electrical Engineering, University of Sheffield, Sheffield S1 3JD, U.K.}
\address[label5]{Department of Electronic and Electrical Engineering, University College London, London, WC1H 0AH, U.K.}






\begin{abstract}
A hole spin is a potential solid-state q-bit, that may be more robust against nuclear spin induced dephasing than an electron spin.
Here we propose and demonstrate the sequential preparation, control and detection of a single hole spin trapped on a self-assembled InGaAs/GaAs quantum dot. The dot is embedded in a photodiode structure under an applied electric-field. Fast, triggered, initialization of a hole spin is achieved by creating a spin-polarized electron-hole pair with a picosecond laser pulse, and in an applied electric-field, waiting for the electron to tunnel leaving a spin-polarized hole. Detection of the hole spin with picosecond time resolution is achieved a second picosecond laser pulse to probe the positive trion transition, where a trion is created conditional on the hole spin to be detected as a change in photocurrent. Finally, using this setup we observe a Rabi rotation of the hole-trion transition that is conditional on the hole spin, which for a pulse-area of $2\pi$ can be used to impart a phase-shift of $\pi$ between the hole spin states, a non-general manipulation of the hole spin.
\end{abstract}


\end{frontmatter}


\section{Introduction}
\label{Intro}

The difference between the on and off states of a typical memory cell in a modern microprocessor is a few hundred electrons.
Under current projections this will reach the limit of one electron in the next few decades. What then? One option under active consideration is a shift from
electronic devices that function classically to those with a quantum functionality, processing quantum bits. For electronic applications, silicon would be the desired material system, but silicon does not provide the optical functionality essential to so many prospective applications. This is where III-V materials step in.
For example, quantum key distribution kits are already commercially available, but can one build a quantum telecom network? Alternatively, linear optics approaches to quantum computing are some of the most advanced. But photon-photon interactions are weak, making deterministic gates impossible to implement without devices with single photon optical nonlinearities. Also, photonic q-bits travel at the speed of light making it difficult to store information at a fixed location.

Atoms have all the quantum properties required to build a q-bit. Atom-like properties can be replicated in semiconductors using quantum dots. The dots trap electrons in discrete energy-levels, energetically isolated from the semiconductor environment, resulting in excitonic (electron-hole pair) transitions with an atom-like light-matter interaction that can be used to control the excitonic and spin states of the dots. Many signatures of an atom-like light-matter interaction have been observed, for example Rabi rotations \cite{Stievater_prl,Zrenner_nat}, Ramsey interference \cite{Bonadeo_sci,Stufler_prl}, Autler-Townes doublet \cite{Kamada_prl,Xu_sci}, and Mollow triplet \cite{Flagg_nphys}. The strong excitonic optical nonlinearity is a result of a short radiative lifetime (0.4-1 ns) \cite{Borri_prl}, limiting the potential use of excitons as q-bits. On the other hand, the spin of a carrier trapped in a dot should be a robust q-bit \cite{Imamoglu_prl}.  Spin relaxation times ($T_1$) of 20-ms, and 0.27-ms have been measured for electron \cite{Kroutvar_nat} and hole \cite{Heiss_prb1} spins respectively, and coherence times ($T_2$) in excess of 3-$\mu$s have been reported for electron spins \cite{Greilich_sci1}. Using the latest growth, and fabrication techniques it is possible to control properties of the dots such as emission wavelength \cite{Rastelli_apl}, providing access to telecom wavelengths \cite{Takaghi_oe},  and positioning \cite{Heidemeyer_prl}. The ability to grow electronically coupled quantum dots \cite{Krenner_prl} offers the possibility of building a few q-bit register. Whilst, recent demonstrations of non-classical light sources \cite{Stevenson_nat,Akimov_prl}, and dot in cavity structures with strong single photon optical nonlinearities \cite{Vuk_nat} are examples of potential quantum devices based on semiconductor quantum dots.

An important capability is the sequential optical initialization, control and read-out of a single spin. Recently, there have been a number of breakthroughs. Continuous-wave pumping schemes for the high-fidelity preparation of an electron \cite{Atature_sci}, and hole \cite{Gerardot_nat} spin have been demonstrated, not only in a Faraday geometry B-field, but also for electron spins in the Voigt geometry B-field needed for optical control \cite{Xu_prl,Kim_prl}. Recently, partial \cite{Berezovsky_sci,Ramsay_prl}, and full \cite{Press_nat} optical control of a single electron spin has been demonstrated.
Full optical control of a single spin is achieved when both the occupation, and relative phase of the spin states can be fully controlled, or in other words, from a well defined initial state, any arbitrary spin state can be prepared. In ref. \cite{Mikkelsen_nphys} the precession of a single electron spin in a GaAs interface dot was observed using a time-resolved Kerr-rotation technique with nanosecond time resolution, later a phase-shift in excess of $\pi/2$ was demonstrated \cite{Berezovsky_sci}. In ref. \cite{Press_nat} full optical control on an initialized electron spin in an InAs dot was demonstrated, where a CW laser was used to both initialize and read-out the spin state, Press {\it et~al} demonstrated both Rabi oscillations, and Ramsey interference of the spin states, showing q-bit rotations about two axes.

In this article, we propose and demonstrate the sequential optical preparation, control, and detection of a single hole spin trapped on an InGaAs/GaAs quantum dot \cite{Ramsay_prl}. Using a dot embedded in an n-i-Schottky diode structure, hole spin initialization is achieved by creating a spin-polarized electron-hole pair with near unit probability using a picosecond laser pulse with a pulse-area of $\pi$, and waiting for the electron to tunnel leaving a spin-polarized hole. Detection of the hole spin is achieved with  picosecond resolution using a second pulse resonant with the positive trion transition to convert the hole spin to a charge degree of freedom detectable as a change in photocurrent. Using this setup we observe a Rabi rotation of the positive trion transition, that is conditional on the hole spin, which for a pulse-area of $2\pi$ can be used to impart a phase-shift of $\pi$ between the hole spin states, a non-general single q-bit operation acting on the hole spin.

\section{Experimental details}

\subsection{Device: n-i-Schottky diode}

The device consists of low density undoped InGaAs/GaAs dots embedded in the intrinsic region of an n-i-Schottky diode structure.
The wafer consists of a GaAs
substrate with the following layers deposited by molecular beam
epitaxy: 50-nm n$^+$ doped GaAs, 25-nm i-GaAs spacer, a single
layer of low density InGaAs dots (30-60 $\mathrm{\mu m^{-2}}$), a
further 125-nm i-GaAs spacer, a 75-nm
$\mathrm{Al_{0.3}Ga_{0.7}As}$ blocking barrier, and a 5-nm i-GaAs
capping layer. The wafer is processed into ($400\times
200~\mathrm{\mu m}$) mesas, with a semi-transparent titanium
top-contact, and an aluminium shadow mask, patterned with 400-nm
apertures using e-beam lithography to spatially isolate single
dots. At low temperatures ($T\sim 10~\mathrm{K}$), the  dark
current of the photodiode in reverse bias is instrument limited
for all voltages of interest: ($\Delta I_{\mathrm{rms}} <
50~\mathrm{fA}, ~I_{\mathrm{offset}}< 200~\mathrm{fA}$, for at
least $\mathrm{V_{reverse}}< 5~\mathrm{V}$).

All of the data presented here is from a single dot, with the neutral exciton emitting at about 1.302 eV.

\subsection{Setup}

The photodiode sits in a cold-finger cryostat at a temperature of about 10-15~K, and is connected to a
low noise IV measurement circuit for photocurrent detection. The dot is excited by a sequence of picosecond laser pulses
derived from a single 150-fs laser pulse from a Ti:sapphire laser with a 76-MHz repetition rate.
The beam is split in two, and each arm passed through an independent pulse-shaper.
The pulse-shapers consist of a 4F zero-dispersion compensator \cite{Weiner_rsi}, with a motorized slit in the masking plane to carve a spectrally narrow ($FWHM=0.2~\mathrm{meV}$ laser pulse with a Gaussian shape from the input pulse.
The pulses are then recombined at a beamsplitter, and focused onto the dot using a long working distance microscope objective. This setup gives independent control of the carrier-frequency, polarization, spectral-width, and pulse-area of each pulse.

We use a photocurrent detection technique \cite{Zrenner_nat}. When a laser resonantly excites excitonic transitions a photon can be absorbed
creating an additional electron-hole pair in the dot. Due to an applied electric-field the carriers tunnel from the dot, and are detected as a photocurrent. The maximum signal corresponding to a single exciton is one electron per measurement cycle, corresponding to 12.18 pA for a repetition rate of 76-MHz. In practice the photocurrent detected for each exciton species is a function of gate voltage, and depends on two factors: the competition between radiative decay and electron tunneling, and a hole tunneling rate that may be slow in comparison to the repetition rate of the laser \cite{Kolodka_prb}. Hence the change in photocurrent detected for a trion, requiring two holes to tunnel, is less than the ideal one electron per pulse. In these experiments the device is intentionally biased for a relatively fast (few ns) hole tunneling rate to achieve a detectable signal. In addition to the photocurrent arising from the dot, there is a background signal proportional to the incident power. This is attributed to absorption of scattered light by other dots in the same mesa that are connected electrically in parallel \cite{Stufler_prb}. This background has been subtracted from all data.

\section{Principle of operation: preparation and detection of single spin}

\begin{figure}
\begin{center}
\includegraphics[scale=0.6]{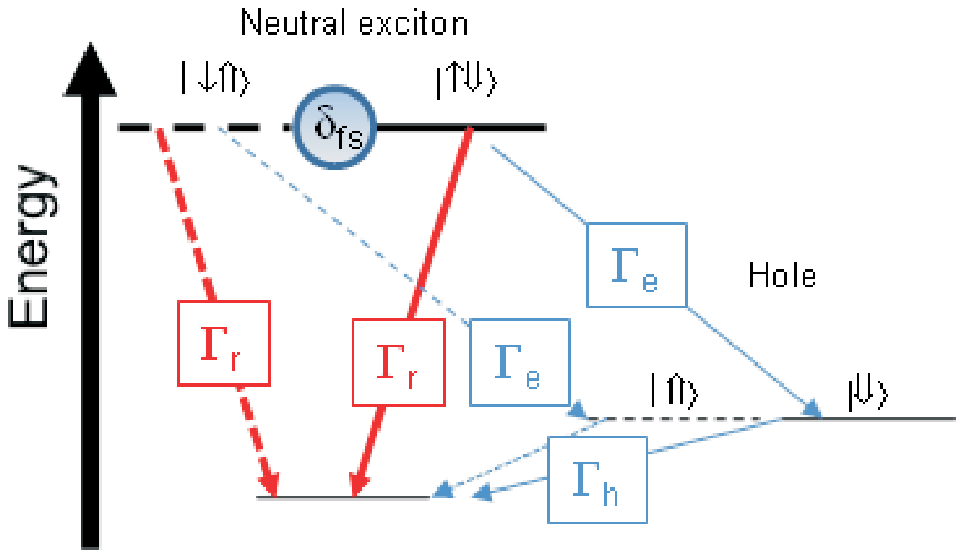}
\end{center}
\caption{ Principle of operation. {\it Preparation} (1) Initially the dot is empty. The first $\sigma_-$-polarized $\pi$-pulse
creates an electron-hole $\vert 0\rangle\rightarrow\mid \uparrow\Downarrow\rangle$ pair with hole spin down $\Downarrow$. (2) The electron tunnels
from the dot leaving a spin-polarized heavy-hole. To prevent the exciton spin precession from erasing the spin information, the device is biased for an electron tunneling rate that is fast compared with the fine-structure splitting. {\it Detection} (3) Due to Pauli-blocking, in the case of a $\sigma_+$-polarized $\pi$-pulse exciting the $\mid\Downarrow\rangle\rightarrow\mid\downarrow\Uparrow\Downarrow\rangle$ transition, a trion is only created if the hole spin-down state is occupied. (4) The electron of the trion tunnels from the dot. The hole spin-down (up) states are mapped to the +2e (+e) charge states, to be detected as a change in photocurrent when the holes tunnel much later.
}\label{fig:POP}
\end{figure}

\subsection{Spin Initialization by exciton Ionization}

To prepare a spin-down heavy-hole $m_j=-3/2$, the photodiode is biased so that the  electron tunneling time is short compared with the fine-structure splitting ($\Gamma_e^{-1}=45\pm 3~\mathrm{ps}\ll 2\pi/\delta_{fs}=(265\pm 10 ~\mathrm{ps})$). A $\sigma_-$ circularly polarized picosecond laser pulse resonantly excites the neutral exciton transition $\vert 0\rangle-\mid\uparrow\Downarrow\rangle$ with a pulse-area of $\pi$. This creates a spin-polarized neutral exciton with near unit probability, and when the electron tunnels from the dot the remaining hole is spin-down $\mid\Downarrow\rangle$. A more detailed description of the spin initialization will now be presented.

The first part of figure \ref{fig:POP} shows the energy-level diagram of a neutral bright exciton, there are two spin states: ($\mid \downarrow\Uparrow\rangle, \mid\uparrow\Downarrow\rangle$), where $\uparrow,\downarrow, (\Uparrow,\Downarrow)$ indicate a ground-state electron (heavy-hole) with  spin $m_s=\pm 1/2, m_j=\pm 3/2$ respectively. Since electric-dipole transitions have a $m_J=\pm 1$ angular momentum selection rule, only the bright-excitons with $m_J=\pm 1$ are optically active.

Due to a small anisotropy, the bright excitons are coupled by the electron-hole exchange interaction to form energy eigen-states that are linear combinations of the exciton spin: $[\vert\pm\rangle=(\mid\downarrow\Uparrow\rangle\pm\mid\uparrow\Downarrow\rangle)/\sqrt{2}]$. The two exciton states and  the crystal ground-state form a 3-level v-transition with linearly polarized selection rules. For a $\sigma_-$-polarized picosecond pulse, with a time-duration that is short compared with the period of the fine-structure beat, the 3-level transition reduces to a 2-level $\vert 0\rangle\leftrightarrow\mid\uparrow\Downarrow\rangle$ transition \cite{Wang_prl,Boyle_prb}. Resonant excitation drives a Rabi oscillation between the $\vert 0\rangle$ and $\mid \uparrow\Downarrow\rangle$ states, at a Rabi frequency $\Omega_R(t)$. A spin-polarized bright exciton $\mid\uparrow\Downarrow\rangle$ is created with near unit probability using a preparation pulse with a pulse-area of $\Theta =\int \Omega_R(t)dt=\pi$.

Under the applied electric-field the electron tunnels from the dot at an unknown time $T$ after the arrival of the preparation pulse, ionizing the dot, where the time $T$ is on the order of the electron tunneling time $\Gamma_e^{-1}$. The initial exciton spin state is a linear superposition of the energy eigenstates $\vert\pm\rangle$, which beat with a frequency equal to the fine-structure splitting. This causes the exciton spin to precess, which acts to erase the spin information. However, if the electron tunneling time is short compared with the period of the fine-structure beat, as is the case in our work, the change in the exciton spin will be small.

\subsection{Photocurrent detection of spin by optical spin to charge conversion}

The energy-level diagram for the heavy-hole/positive trion 4-level system is shown in the middle section of figure \ref{fig:POP}. For zero applied B-field the hole states $\mid\Uparrow\rangle,\mid\Downarrow\rangle$ are degenerate, as are the positive trion states $\mid\uparrow\Uparrow\Downarrow\rangle,\mid\downarrow\Uparrow\Downarrow\rangle$. A spectrally narrow (FWHM=0.2 meV) laser pulse is used to resonantly excite the trion transition only. Absorption of a $\sigma_+$ circularly polarized laser pulse would create an electron-hole pair ($\downarrow\Uparrow$). However, the creation of two holes of the same spin is forbidden by the Pauli exclusion principle, so the absorption only occurs if the dot is occupied by a spin-down hole $\mid\Downarrow\rangle$. If a trion is created, the electron will tunnel from the dot leaving the dot charged with two holes. If a trion is not created, the dot remains singly charged. In other words, the circular polarization of the detection pulse of pulse-area $\pi$ can be used to select a hole-spin state to map to a trion state, and then electron tunneling converts this information on the hole spin state to a charge degree of freedom, that is electrically detected as a change in photocurrent. A spin to charge conversion scheme suitable for photoluminescence detection was recently proposed by Heiss {\it et~al} \cite{Heiss_prb2}.

\section{Experimental demonstration of preparation and detection}

\begin{figure}
\begin{center}
\includegraphics[scale=0.6]{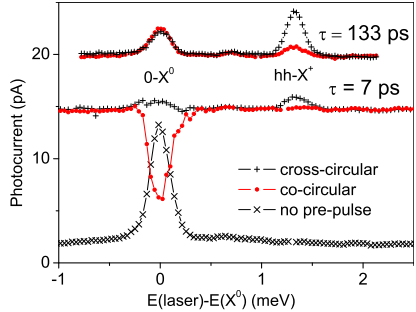}
\end{center}
\caption{ Preparation and detection of a single hole spin. Photocurrent versus laser detuning of the detection pulse of pulse-area $\pi$.
(lower) For single pulse excitation, a single peak is observed for the neutral exciton transition.
(middle and upper) Two color photocurrent spectra, offset for clarity. The key feature is the additional peak due to the $h-X^+$ transition
observed predominantly for cross-circular excitation.
}\label{fig:spec}
\end{figure}

Figure \ref{fig:spec} presents two-color photocurrent spectra showing the preparation and detection of a single hole spin, at a reverse gate voltage of 0.8~V.
In the first experiment, a single $\sigma_-$ polarized pulse excites the dot, with a pulse-area of $\pi$. The photocurrent is measured versus the laser detuning $\hbar\delta_i=E_i-E(X^0), (i = 1,2)$ with respect to the neutral exciton transition. A single peak is observed, as seen in the lower part of fig. \ref{fig:spec}. The lineshape is dominated by the Gaussian spectral pulse-shape of the excitation pulse with a FWHM= 0.2 meV. The pulse-area is calibrated by measuring a Rabi rotation of the neutral exciton transition \cite{Kolodka_prb}.

In the second set of experiments, two circularly polarized pulses with a pulse-area of $\pi$ excite the dot. The first pulse, termed the preparation pulse, excites the neutral exciton transition on resonance ($\delta_1=0$) with $\sigma_-$ polarization, creating a spin-polarized exciton $\mid \uparrow\Downarrow\rangle$. The second pulse, labeled the detection pulse, excites the dot at a time-delay $\tau$ after the preparation pulse. No offset has been subtracted from the middle trace of fig. \ref{fig:spec}. When the detection pulse is off-resonance from any transitions it is not absorbed, and a photocurrent corresponding to one neutral exciton, and some additional background is detected. For the middle trace, the time-delay $\tau=7~\mathrm{ps}$ is short compared with the electron tunneling time, and the dot has a significant exciton population. For co-circular excitation, a dip in the photocurrent is observed on-resonance with the $0-X^0$ transition. Here, the two $\pi$-pulses act like a single $2\pi$ pulse leading to minimal creation of the exciton. For cross-circular excitation, a small increase in photocurrent is observed on-resonance with the $0-X^0$ transition. Here absorption is suppressed, since the photon-energy required to create a biexciton is shifted by the biexciton binding energy of 1.9 meV. Most importantly, an additional peak is observed corresponding to the hole to positive trion $h-X^+$ transition. For a time-delay of $\tau =133~\mathrm{ps}$ that is long compared with the electron tunneling time (upper traces), the amplitude of the trion peak has increased, indicating that the electron has tunneled from the dot leaving the spin-polarized hole. A much smaller $h-X^+$ peak is also observed for co-circular excitation, indicating that the spin preparation is not perfect, and some spin scattering has occurred. At large time-delays a polarization insensitive peak is observed at the $0-X^0$ transition, indicating that there is a small probability that the dot is unoccupied at time $\tau$. The two-color photocurrent spectra show that it is possible to prepare a single hole spin, and to detect that spin state.

Two-color photocurrent spectroscopy can also be used to detect the exciton-biexciton transition, providing access to the physically rich exciton-biexciton system. For further details the reader is referred to refs. \cite{Boyle_prb,Boyle_man}.

\begin{figure}
\begin{center}
\includegraphics[scale=0.6]{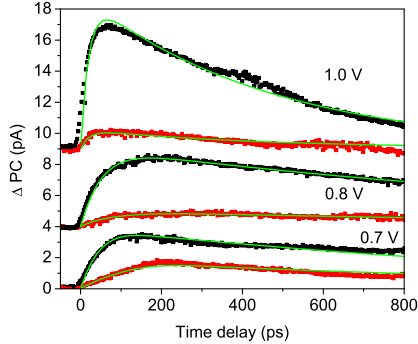}
\end{center}
\caption{Pump-probe style measurements of the hole spin, at various voltages,  co-circular (red), and cross-circular (black) excitation.}\label{fig:time}
\end{figure}

Pump-probe  measurements of the hole spin can be made. Time-resolved measurements of the hole-spin are presented for various gate voltages in fig. \ref{fig:time}.
Here a change in photocurrent $\Delta PC(\tau)=PC(\tau)-PC(-\infty)$, proportional to the occupation of the targeted hole spin state at time $\tau$, is plotted versus the time-delay between the preparation and detection pulses with pulse-areas of $\pi$. Once again, the $\sigma_-$-polarized preparation pulse is resonant with the neutral exciton $0-X$ transition, and creates a spin polarized neutral exciton. The detection pulse is on-resonance with the positive trion $h-X^+$ transition, and is circularly polarized. The black traces present data for cross-circular excitation to detect the desired spin state, whilst the red traces show co-circular excitation to measure the undesired spin state. For short time-delays the photocurrent increases exponentially as the electron tunnels from the dot, reaches a maximum, and then decays exponentially due to the hole tunneling from the dot. The rate of the initial rise is slower for co-circular excitation, because here the timescale is determined by the slow fine-structure beat. As the gate voltage is increased, both the initial rise, and later decay of the trace speeds up due to the increased tunneling rates. More importantly the proportion of the signal in the desired spin state increases as the electron tunneling rate begins to dominate over the fine-structure beat. These are time-resolved measurements of a single hole spin with a time-resolution limited by the picosecond time duration of the detection pulse.

\section{Measurements of exciton fine structure beats}

\begin{figure}
\begin{center}
\includegraphics[scale=0.6]{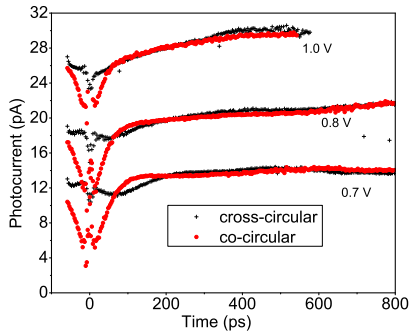}
\end{center}
\caption{Inversion recovery measurements of the neutral exciton transition for various voltages for co and cross-polarized excitation. The photocurrent is measured as a function of the time-delay between two $\pi$ pulses, and is proportional to the population inversion between the exciton spin selected by the polarization of the second pulse, and the crystal ground-state. }\label{fig:X0}
\end{figure}

To test our understanding of the hole initialization, we need to know the fine-structure splitting and the electron tunneling rates. To this end, time-resolved measurements of the exciton fine structure beats are made using a polarization-resolved inversion recovery technique \cite{Kolodka_prb}. The data are presented in fig \ref{fig:X0}, where the photocurrent is measured as a function of the time-delay between two circularly polarized $\pi$-pulses, termed the pump and probe, that resonantly excite the neutral exciton transition. The measured change in photocurrent is proportional to the population inversion between the exciton spin state $x_t(\tau)$ selected by the polarization of the probe pulse, and the crystal ground-state $x_0(\tau)$: $PC_t(\tau)\propto 1-(x_t(\tau)-x_0(\tau))$. Initially the dot is in the crystal ground-state. The  pump performs a Rabi-rotation through an angle of $\pi$ creating a spin-polarized exciton, which evolves for a time $\tau$ before the arrival of the probe. The polarization of the probe selects one of the exciton transitions, and drives a Rabi rotation through an angle $\pi$ inverting the populations of the selected exciton spin state and the crystal ground state, resulting in a change in photocurrent proportional to this population inversion.

Figure \ref{fig:X0} presents data for inversion recovery measurements of the neutral exciton transition for both co, and cross circular polarizations.
In the case of co-circular excitation, at zero time-delay the pulses interfere resulting in rather noisy data. When the pulses no longer overlap in time the photocurrent is initially low, since the effect of two $\pi$ pulses is equivalent to one $2\pi$-pulse with the net effect that no exciton is created by the pulse-pair. There is an exponential saturation in the photocurrent as the electron tunnels from the dot, followed by a slow exponential saturation as the hole tunnels from the dot, since the second pulse can only be absorbed if the dot is in the crystal ground-state following a hole tunneling event. In addition there is an oscillation due to the exciton fine-structure beat. The cross-circular excitation probes the orthogonal exciton spin, and a fine-structure beat in anti-phase with the co-circular case is observed. As the voltage is increased the decay rates speed up due to the increased tunneling rates.
In the next section we will present a model for both the data in fig. \ref{fig:time}, and fig. \ref{fig:X0}.



\section{Model of spin preparation}

\begin{figure}
\begin{center}
\includegraphics[scale=0.6]{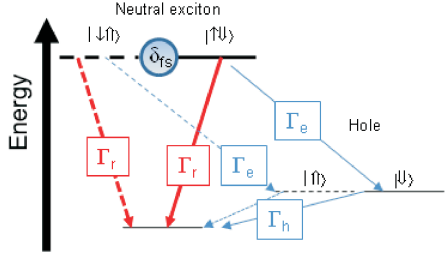}
\end{center}
\caption{ Energy-level diagram of rate-equation model for spin preparation with
electron tunneling rate $\Gamma_e$, Hole tunneling rate $\Gamma_h$, radiative recombination rate $\Gamma_r$
and a fine-structure beat frequency $\delta_{fs}$.
}\label{fig:model}
\end{figure}

\begin{figure}
\begin{center}
\includegraphics[scale=0.6]{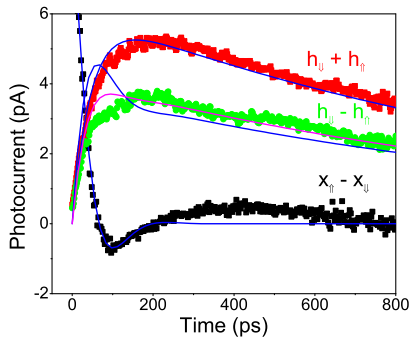}
\end{center}
\caption{ Comparison of experiment and model at gate voltage of 0.8~V. (black) Difference in the occupations of the exciton spin states: $(x_{\Downarrow}-x_{\Uparrow})$. (red) Total hole population: $h_{\Downarrow}+h_{\Uparrow}$. (green) Difference in occupations of hole spin states: $h_{\Downarrow}-h_{\Uparrow}$. (blue) Fits to data using model A. (magenta) Fits to data using model B.
}\label{fig:fits}
\end{figure}

To model the hole spin preparation, and the time-resolved traces we consider a simple rate equation model that is shown schematically in fig. \ref{fig:model}. For simplicity, we assume that the preparation pulse is a $\delta$-pulse that creates a spin-polarized exciton at time zero, and there are no spin-flip processes. First we consider the evolution of the exciton spin, where the occupations of the exciton spin state are $x_{\Downarrow},x_{\Uparrow}$ where the hole spin is used to label the states:

\begin{eqnarray}
\dot{x}_{\Downarrow}+\dot{x}_{\Uparrow}= -\Gamma_X (x_{\Downarrow}+x_{\Uparrow}) \\
\ddot{x}_{\Downarrow}-\ddot{x}_{\Uparrow}= -\Gamma_X (\dot{x}_{\Downarrow}-\dot{x}_{\Uparrow})+\delta_{fs}^2 (x_{\Downarrow}-x_{\Uparrow})
\end{eqnarray}

\noindent where $\Gamma_e \gg \Gamma_h$ are the electron and hole tunneling rates, $\Gamma_r$ the radiative recombination rate, and $\delta_{fs}$ is the frequency of the fine-structure splitting, and $\Gamma_X=\Gamma_e+\Gamma_r$. The exciton spin is time-resolved in the inversion recovery measurements presented in fig. \ref{fig:X0}. The difference in photocurrent between measurements made with  co and cross-circular excitation is proportional to the exciton spin inversion $(x_{\Downarrow}-x_{\Uparrow})$, and for a representative gate-voltage of 0.8~V the strongly damped exciton fine-structure beat is presented as the black trace in fig. \ref{fig:fits}. The blue trace shows a fit to the exciton spin inversion using: $x_{\Downarrow}-x_{\Uparrow}=Ae^{-\Gamma_X\tau}\mathrm{cos}(\delta_{fs}\tau)$, and values for the exciton decay rate, and the fine-structure are determined, with $\Gamma_X^{-1}=45\pm 3~\mathrm{ps}$, and $2\pi/\delta_{fs}=265\pm 10~\mathrm{ps}$.

When an electron tunnels from the dot, the exciton is ionized leaving a hole. Therefore the evolution of the total hole population is described by:

\begin{equation}
\dot{h}_{\Downarrow}+\dot{h}_{\Uparrow}=+\Gamma_e (x_{\Downarrow}+x_{\Uparrow}) - \Gamma_h(h_{\Downarrow}+h_{\Uparrow})
\end{equation}

\noindent where $h_{\Downarrow},h_{\Uparrow}$ are the occupations of the hole spin states. The total hole spin population is proportional to the sum of the photocurrents measured in the co and cross circular measurements shown in fig. \ref{fig:time}, and plotted as the red trace in fig. \ref{fig:fits}. The blue trace is a fit to: $(h_{\Downarrow}+h_{\Uparrow}) = B(e^{-\Gamma_h\tau}-e^{-\Gamma_X\tau})$, with $\Gamma_X^{-1}= 45~\mathrm{ps}$ fixed by the exciton spin inversion measurements, and a value of $\Gamma_h^{-1} = 1.3~\mathrm{ns}$ found from the fit.

\subsection{What happens to the hole spin when the exciton is ionized?}

Here we will consider two different models to describe what happens to the hole spin when the electron tunnels from the dot.

\subsubsection{Model A}
In `model A' when the electron tunnels from the dot the electron spin is traced out, in which case the exciton spin is mapped directly to the hole spin state. Here the hole spin inversion evolves as:

\begin{equation}
\dot{h}_{\Downarrow}-\dot{h}_{\Uparrow}=+(2p-1)\Gamma_e (x_{\Downarrow}-x_{\Uparrow}) -\Gamma_h(h_{\Downarrow}-h_{\Uparrow})
\end{equation}

where $p\leq 1$ is the probability that when the electron tunnels from the dot, the hole spin is not flipped. To compare this model to the data, the hole spin inversion, which  is proportional to the difference in photocurrent measured for the co and cross-circular polarized measurements is calculated using the data in fig. \ref{fig:time}, and plotted as the green trace presented in fig, \ref{fig:fits}.
 The blue line shows a fit to data using the values measured for $\Gamma_X,\Gamma_h,\delta_{fs}$, where the amplitude of the signal is the only fitting parameter. As can be seen in fig. \ref{fig:fits}, `model A' does not describe the data very well. In `model A' the hole spin inversion contains some information on the exciton spin coherence, and as a result overshoots the maximum, whereas in the data the oscillation is suppressed. Another failing of the model is that it underestimates the measured contrast $\mathcal{C}$ of the spin preparation, this aspect will be discussed later.

\subsubsection{Model B}
The quality of the spin preparation improves with increased electron tunneling rate, indicating that the depolarization of the hole spin is caused by the electron-hole exchange interaction. However, the authors do not fully understand what happens to the hole spin when the electron tunnels from the dot, when after some time the electron-hole exchange interaction has entangled the electron and hole spin. However, whatever does occur acts to improve the spin preparation. This could be a subtle question. If the environment `measures' the electron spin, this could affect the outcome of the spin preparation in a non-trivial fashion. To mimic this behavior we try a phenomenological `model B', where if the exciton is in the hole-spin down $\Downarrow$-state no spin-flip has occurred and the exciton relaxes to the hole spin down state. If however the exciton is in the hole-spin up $\Uparrow$-state a spin-flip has occurred, and the hole spin is randomized. This model assumes that the electron tunneling is too fast for the exciton spin to undergo a full cycle of the fine-structure beat, which is the case in our experiments. The model reads as:

\begin{equation}
\dot{h}_{\Downarrow}-\dot{h}_{\Uparrow}=+(2p-1)\Gamma_e x_{\Downarrow} -\Gamma_h(h_{\Downarrow}-h_{\Uparrow})
\end{equation}

A fit to `model B' is presented in fig. \ref{fig:fits} as a magenta line, and gives a better fit to the data. Here the amplitude is the only fitting parameter, and the  measured values for $\Gamma_X,\Gamma_h,\delta_{fs}$ were used.

\subsection{Fidelity of spin preparation}

\begin{figure}
\begin{center}
\includegraphics[scale=0.6]{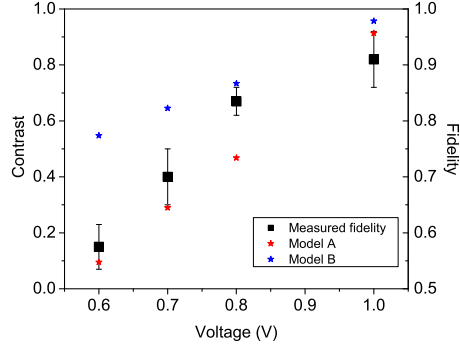}
\end{center}
\caption{ Improved fidelity of spin preparation at increased reverse bias. (black squares) Contrast of spin preparation $\mathcal{C}$ measured using data in fig. \ref{fig:time}. (red stars) Model A. (blue stars) Model B.
}\label{fig:fidelity}
\end{figure}

To evaluate the quality of the spin preparation we define a contrast of the hole spin preparation $\mathcal{C}$ as the ratio:

\begin{equation}
\mathcal{C}=\lim_{\Gamma_X\tau\gg 1} \frac{p_{\Downarrow}-p_{\Uparrow}}{p_{\Downarrow}+p_{\Uparrow}}
\end{equation}

Figure \ref{fig:fidelity} presents the voltage dependence of the contrast, and the fidelity, defined here as ($\mathcal{F}=(\mathcal{C}+1)/2$).
The black squares show the contrast measured using the data in fig.\ref{fig:time}, and for comparison with the model
a contrast is calculated for `model A' (red circles)of:

\begin{equation}
\mathcal{C}_A= \frac{2p-1}{1+(\frac{\delta_{fs}}{\Gamma_X-\Gamma_h})^2}
\end{equation}

\noindent and for `model B' (blue triangles) of:

\begin{equation}
\mathcal{C}_B= \frac{2p-1}{2}[1+\frac{1}{1+(\frac{\delta_{fs}}{\Gamma_X-\Gamma_h})^2}]
\end{equation}

 The theoretical contrasts are calculated for the ideal case $p=1$ using measured values for the fine-structure splitting of $\delta_{fs}=\frac{2\pi}{265\mathrm{ps}}$, and electron tunneling times of  $\Gamma_e^{-1}=\{130, 66, 45, 13\}~\mathrm{ps}$.
The radiative recombination, and hole tunneling rates are neglected. At low reverse bias the contrast is closer to that predicted by `model A', and at higher bias the contrast is closer to that predicted for `model B'.

To conclude this section. In the experiments the contrast of the hole spin preparation improves with increased electron tunneling rate indicating that the depolarization of the hole spin is a result of the exciton spin precession. However, what happens to the hole spin when an exciton with an entanglement between the electron and hole spins  is ionized is not yet clear, and warrants further detailed investigation.

\section{Routes to improved fidelity of spin preparation}

The experiments presented here are for zero applied B-field. In a Faraday geometry, with a B-field aligned along the growth direction, high fidelity spin preparation should be possible. Firstly the energy eigenstates of the exciton, and the hole, will be mostly aligned along the B-field, and the exciton spin created by the preparation pulse will not precess. Secondly the Zeeman splitting will introduce energy selectivity to the excitation making the polarization of the preparation pulse less critical. Of more importance, is the quality of the spin initialization in the Voigt geometry, with a B-field aligned parallel to the growth direction, since this is the geometry needed for full optical control. Using circularly polarized excitation, the quality of the spin preparation is likely to deteriorate as the Zeeman energy effectively increases the fine-structure splitting. One possibility, is to use a linearly polarized preparation pulse to create an exciton with a spin aligned along the B-field, in an energy eigenstate that will not precess, and that may then tunnel to the corresponding hole spin state which would also be an eigenstate.

The fidelity of the preparation could also be improved by eliminating the fine-structure splitting. For some dots this can be achieved by applying a Voigt geometry B-field \cite{Stevenson_prb1}, thermal annealing of the dot \cite{Young_prb,Ellis_apl}, or possibly by applying a lateral electric-field to the dot \cite{lateral_apl}.

\section{Conditional Rabi rotation of positive trion}

\begin{figure}
\begin{center}
\includegraphics[scale=0.6]{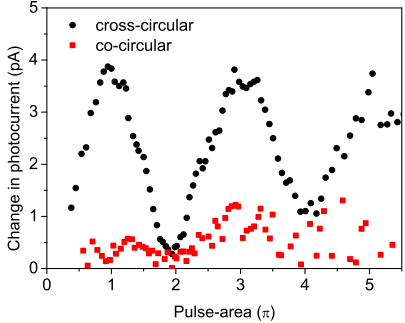}
\end{center}
\caption{ Rabi rotation of positive trion transition, conditional on the hole spin. The hole spin is prepared using a $\sigma_-$ polarized pulse.
The change in photocurrent versus the pulse-area of a second control pulse on resonance with the $h-X^+$ transition is plotted. The Rabi rotation is only observed for cross-circular excitation.
}\label{fig:RR1}
\end{figure}

All proposals for the coherent optical control of a spin rely on the trion transitions having an atom-like light-matter interaction.
The most convincing demonstration of this property is the observation of a Rabi rotation, conditional on the spin of the carrier. Here we demonstrate this for the positive trion transition of an InAs dot.  Figure \ref{fig:RR1} presents a measurement of a Rabi rotation of the hole to positive trion $h-X^+$ transition that is conditional on the hole spin. Two pulses excite the dot. The first is a $\sigma_-$-polarized preparation pulse of pulse-area $\pi$ on-resonance with the neutral exciton $0-X^0$ transition, and prepares a spin-polarized exciton. The second pulse, now termed the control, excites the dot on-resonance with the hole-trion $h-X^+$ transition, at a time-delay  $\tau =133~\mathrm{ps}$ that is long compared with the electron tunneling time. Initially the dot is in a mostly hole spin-down $\mid\Downarrow\rangle$ state. The change in photocurrent, which is proportional to the final occupation of the trion state, is measured as a function of the pulse-area of the control pulse. In the case of cross-circular excitation, a Rabi rotation is observed for rotation angles in excess of $5\pi$, whereas for co-circular the change in photocurrent is much smaller. Thus demonstrating a Rabi rotation of a trion transition conditional on the initialized hole spin in a single InGaAs/GaAs quantum dot.

\begin{figure}
\begin{center}
\includegraphics[scale=0.6]{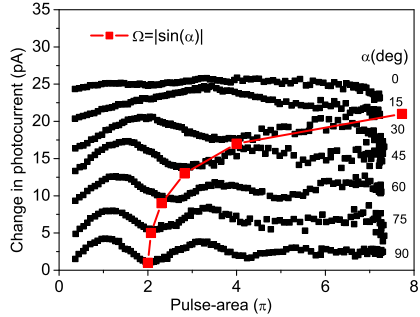}
\end{center}
\caption{ Evidence for the trion transitions acting like two independent 2-level transitions. The hole spin is prepared using a $\sigma_-$ polarized pulse.
The change in photocurrent versus the pulse-area of a second control pulse on resonance with the $h-X^+$ transition is plotted. The polarization of the control is varied as $\hat{e}=\hat{\sigma}_-\mathrm{cos}(\alpha)+\hat{\sigma}_+\mathrm{sin}(\alpha)$. The period of the Rabi rotation changes because only the $\hat{\sigma}_+$ component of the control pulse drives the Rabi rotation. The red-line indicates the $\vert \mathrm{sin}(\alpha)\vert$ dependence of the observed Rabi frequency.
}\label{fig:RR2}
\end{figure}

For zero applied B-field, the 4-level hole-trion system should act as two independent two-level transitions: ($\mid\Downarrow\rangle\leftrightarrow\mid\downarrow\Uparrow\Downarrow\rangle, \mid\Uparrow\rangle\leftrightarrow\mid\uparrow\Uparrow\Downarrow\rangle$), as illustrated in fig. \ref{fig:POP}. To confirm this, we measured the dependence of the Rabi rotation on the polarization of the control pulse. The data is shown in fig. \ref{fig:RR2}. Initially the hole is spin-down $\mid\Downarrow\rangle$. Rabi rotations of the hole-trion $h-X^+$ transition are measure as the polarization of the control pulse is varied: $\hat{e}=\hat{\sigma}_-\mathrm{cos}(\alpha)+\hat{\sigma}_+\mathrm{sin}(\alpha)$. The key observation is that the inverse period of the Rabi rotation is proportional to $\vert\mathrm{sin}(\alpha)\vert$ showing that only the $\sigma_+$ component of the control pulse couples to the $\mid\Downarrow\rangle\leftrightarrow\mid\downarrow\Uparrow\Downarrow\rangle$ transition. Hence, the trion transitions behave as two independent two-level transitions. Recently Kim {\it et~al} have observed the Rabi rotation of a negative trion in InAs dot using a differential transmission technique \cite{Kim_cleo}. Rabi rotations have also been reported for an ensemble of InAs/GaAs dots \cite{Greilich_prl}, and for a p-shell transition where the dot has an unknown charge state \cite{Besombes_prl}.

Here we have demonstrated a Rabi rotation of the positive trion transition that is conditional on the hole spin. Such a manipulation can be used to impart a relative phase-shift of $\pi$ between the hole spin states, a non-general single q-bit operation, as discussed in the next section.

\section{Pauli-z operation}

\begin{figure}
\begin{center}
\includegraphics[scale=0.6]{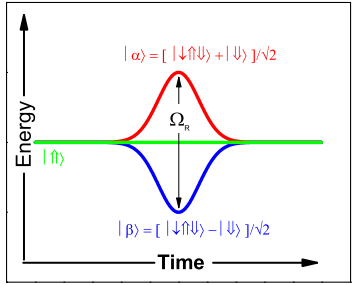}
\end{center}
\caption{ Principle of geometric $\sigma_z$ gate to impart a relative phase-shift of $\pi$ between the hole spin states.
The $\sigma_+$-polarized control pulse couples the $\mid\Downarrow\rangle,\mid\downarrow\Uparrow\Downarrow\rangle$ states, and in the rotating frame of the control laser, the energy eigen-states are: $\mid\Uparrow\rangle, \vert\alpha\rangle=[\mid\Downarrow\rangle+\mid\downarrow\Uparrow\Downarrow\rangle]/\sqrt{2},
\vert\beta\rangle=[\mid\Downarrow\rangle-\mid\downarrow\Uparrow\Downarrow\rangle]/\sqrt{2}$, with energies of $0,\pm\Omega_R/2$ respectively.
The control pulse splits the dressed states $\vert\alpha\rangle$ and $\vert\beta\rangle$ causing the superposition to beat with the Rabi frequency. An initial state of:  $a\mid\Uparrow\rangle+b[\vert\alpha\rangle+\vert\beta\rangle]/\sqrt{2}$ evolves to: $a\mid\Uparrow\rangle+b[e^{-i\Theta/2}\mid\alpha\rangle+e^{i\Theta/2}\mid\beta\rangle]/\sqrt{2}, \Theta=\int\Omega_R(t)dt$. In the case of $\Theta=2\pi$, the state-vector returns to the hole-only subspace, having imparted a relative phase-shift of $\pi$ between the hole states $a\mid\Uparrow\rangle-b\mid\Downarrow\rangle$.
}\label{fig:Pauli}
\end{figure}

One of the set of single q-bit operations needed for fault tolerant quantum information processing is the $\sigma_z$ operator. Where the control pulse imparts a relative phase-shift of $\pi$ between the logical states. A unitary operator described by the Pauli $\sigma_z$ matrix, hence the name. For a q-bit encoded in the $m_j=\pm 3/2$ spin-states of the heavy-hole this can be achieved using a circularly polarized control pulse to drive a Rabi rotation of the trion transition through an angle of $2\pi$ as observed in figs. \ref{fig:RR1}, and \ref{fig:RR2}, and discussed theoretically in refs. \cite{Economou_prl,Nazir_prl}. The principle of operation is illustrated in figure \ref{fig:Pauli}, showing an energy-time graph of the hole-trion states in the rotating frame of the control pulse. The $\sigma_+$ circularly polarized control pulse couples the $\mid\Downarrow\rangle,\mid\downarrow\Uparrow\Downarrow\rangle$ states only. Consequently the energy-eigenstates of the system are the dressed states $\vert\alpha\rangle,\vert\beta\rangle=[\mid\Downarrow\rangle\pm\mid\downarrow\Uparrow\Downarrow\rangle]/\sqrt{2}$, with energies of $\pm\Omega_R(t)/2$ respectively.
If we assume the hole spin is initially in a state: $a\mid\Uparrow\rangle+b\mid\Downarrow\rangle=a\mid\Uparrow\rangle+b[\vert\alpha\rangle+\vert\beta\rangle]/\sqrt{2}$. The control pulse shifts the energy of the dressed states, causing each to  accumulate a relative phase $\Delta\phi$ equal to the integral of the energy-shift $\Delta\phi=\mp\int\Omega_R(t)dt/2$, and with respect to the uncoupled reference state $\mid\Uparrow\rangle$. This causes the hole superposition to evolve to: $a\mid\Uparrow\rangle+b[e^{-i\Theta/2}\mid\alpha\rangle+e^{i\Theta/2}\mid\beta\rangle]/\sqrt{2}$. In the case of a control pulse of pulse-area $2\pi$, the final state-vector is in the hole sub-space only, and a relative phase-shift of $\pi$ is imparted between the hole spin states resulting in a final state of: $a\mid\Uparrow\rangle-b\mid\Downarrow\rangle$. This geometric phase-shift has been observed for a neutral exciton in a GaAs interface quantum dot using a four-wave mixing technique \cite{Patton_prl}, supporting this interpretation.

\section{In conclusion}

We have proposed and demonstrated the sequential preparation, control and detection of a single hole spin trapped on an InGaAs/GaAs quantum dot. Our scheme includes a method for high fidelity, triggered preparation of a single spin without the need for a B-field, and a photocurrent detection method capable of the picosecond time resolution needed for evaluating the performance of coherent control operations. Evidence for a non-general manipulation  of the hole spin is observed as a Rabi rotation of the hole-trion transition that is conditional on the hole spin state.

The use of photocurrent detection gives the advantage of efficient detection, with good signal to noise, enabling relatively fast measurement runs. A typical Rabi rotation measurement, as presented here takes a minute to acquire. The penalty is a hole coherence time  limited by a hole tunneling rate that needs to be faster than the repetition rate of the laser. At the moment, for time-resolved measurements of single quantum dot spins in the coherent control regime, the spin is intentionally decohered to achieve useable signal strengths, compatible with MHz repetition rates. Either by using GaAs interface dots with short spin coherence times \cite{Berezovsky_sci}, or a read-out laser that is the main source of decoherence \cite{Press_nat}. For the future, what is needed is a measurement scheme where the measurement is gated, rather than always on. This is an attractive feature of  the Kerr-rotation method reported in ref.\cite{Berezovsky_sci}. A key advantage of using a photodiode structure is the potential to use a dynamic gate voltage to switch the device between a low tunneling rate regime with long coherence times, and high tunneling rate for efficient detection.

Another important distinction between this and other work is the use of a hole, rather than electron spin. The main source of dephasing for the electron spin, is the interaction with the nuclear spins via the hyperfine interaction \cite{Greilich_sci1}. Since the wavefunction of the heavy-hole has a p-type Bloch-function, the contact hyperfine interaction is zero, and hence the nuclear spin dephasing should be much weaker for holes \cite{Laurent_prl}. Although recently Fischer {\it et~al} \cite{Loss_prb} have pointed out that the hole is not entirely immune to the effects of nuclear spin. At the moment it is not clear if electron or hole spins will make better q-bits, but in principle,  one could apply the same ideas to a p-type Schottky diode to prepare, and detect a single electron spin.

Finally, these experiments are for InGaAs/GaAs self-assembled quantum dots with high optical quality rather than GaAs interface dots.
Recently, Wu {\it et~al} concluded that it was not possible to observe a Rabi rotation of the negative trion in GaAs dots \cite{Wu_prl}. This suggests that for the trion transitions the interactions with the environment may be too strong in GaAs dots for the transitions to be regarded as atom-like.

\section{Acknowledgements}

The authors thank the EPSRC U.K. GR/S76076, and QIPIRC U.K. for funding this work. JBBO acknowledges financial support from CAPES Brazil.
AFAK acknowledges financial support from the University of Cairo.



\end{document}